\documentclass{article}

    \PassOptionsToPackage{numbers, compress}{natbib}

    \usepackage[final]{neurips_2024}

\usepackage[utf8]{inputenc} 
\usepackage[T1]{fontenc}    
\usepackage{hyperref}       
\usepackage{url}            
\usepackage{booktabs}       
\usepackage{amsfonts}       
\usepackage{nicefrac}       
\usepackage{microtype}      
\usepackage{xcolor}         
\usepackage{graphicx}
\usepackage{amsmath}
\usepackage[version=3]{mhchem}
\usepackage[normalem]{ulem}

\title{GFlowNets for Hamiltonian decomposition in groups of compatible operators}

\author{%
    Isaac L. Huidobro-Meezs\\
    Department of Chemistry and Chemical Biology\\
    McMaster University
    \And
    Jun Dai \\
    Mila \& DIRO\\
    Université de Montréal\\
    Montréal, QC, Canada \\
    \And
    Guillaume Rabusseau \\
    Mila \& DIRO \\
    Université de Montréal \\
    Montréal, QC, Canada \\
    \And
    Rodrigo A.~Vargas-Hernández$^{*}$\\
    Department of Chemistry and Chemical Biology\\
    McMaster University\\
    $^{*}$\texttt{vargashr@mcmaster.ca}
}

\begin{document}

\maketitle

\begin{abstract}
Quantum computing presents a promising alternative for the direct simulation of quantum systems with the potential to explore chemical problems beyond the capabilities of classical methods. However, current quantum algorithms are constrained by hardware limitations and the increased number of measurements required to achieve chemical accuracy.
To address the measurement challenge, techniques for grouping commuting and anti-commuting terms, driven by heuristics, have been developed to reduce the number of measurements needed in quantum algorithms on near-term quantum devices. In this work, we propose a probabilistic framework using GFlowNets to group fully (FC) or qubit-wise commuting (QWC) terms within a given Hamiltonian. The significance of this approach is demonstrated by the reduced number of measurements for the found groupings; 51\% and 67\% reduction factors respectively for FC and QWC partitionings with respect to greedy coloring algorithms, highlighting the potential of GFlowNets for future applications in the measurement problem. Furthermore, the flexibility of our algorithm extends its applicability to other resource optimization problems in Hamiltonian simulation, such as circuit design.
\end{abstract}

\section{Introduction}

Quantum computing has gained considerable attention for its potential to solve the electronic structure problem (ESP), a fundamental challenge in computational chemistry and material science~\citep{VQE:review:2019,VQE:review:2020,VQE:review:2020b,VQE:review:2022,VQE-Rev}. However, a major obstacle for quantum algorithms addressing the ESP, particularly on current noisy intermediate-scale quantum (NISQ) devices, is the measurement problem—the large number of measurements required to achieve chemical accuracy~\citep{su2021fault}.
To mitigate this, two main approaches have been developed: (i) grouping commuting/anti-commuting terms and applying factorization/modification techniques to the Hamiltonian~\citep{huggins2021efficient, oumarou2022accelerating,izmaylov2019unitary, Thomson-npj,izmaylov2020JCPmeasurements,izmaylov2021measurements,Nacho2,Nacho_2023,Nachofluidfermionic}, and (ii) partial-tomography protocols inspired by shadow tomography~\citep{huang2020predicting,hadfield2022measurements,garcia2021learning}. A third approach integrates both frameworks~\citep{zhang2023measurement}.
While these methods have made substantial progress in tackling the measurement problem, their reliance on heuristics for optimization or the grouping per-se does not guarantee that the selected grouping minimizes the number of measurements. 

\begin{figure}[ht!]
    \centering
    \includegraphics[scale=0.23]{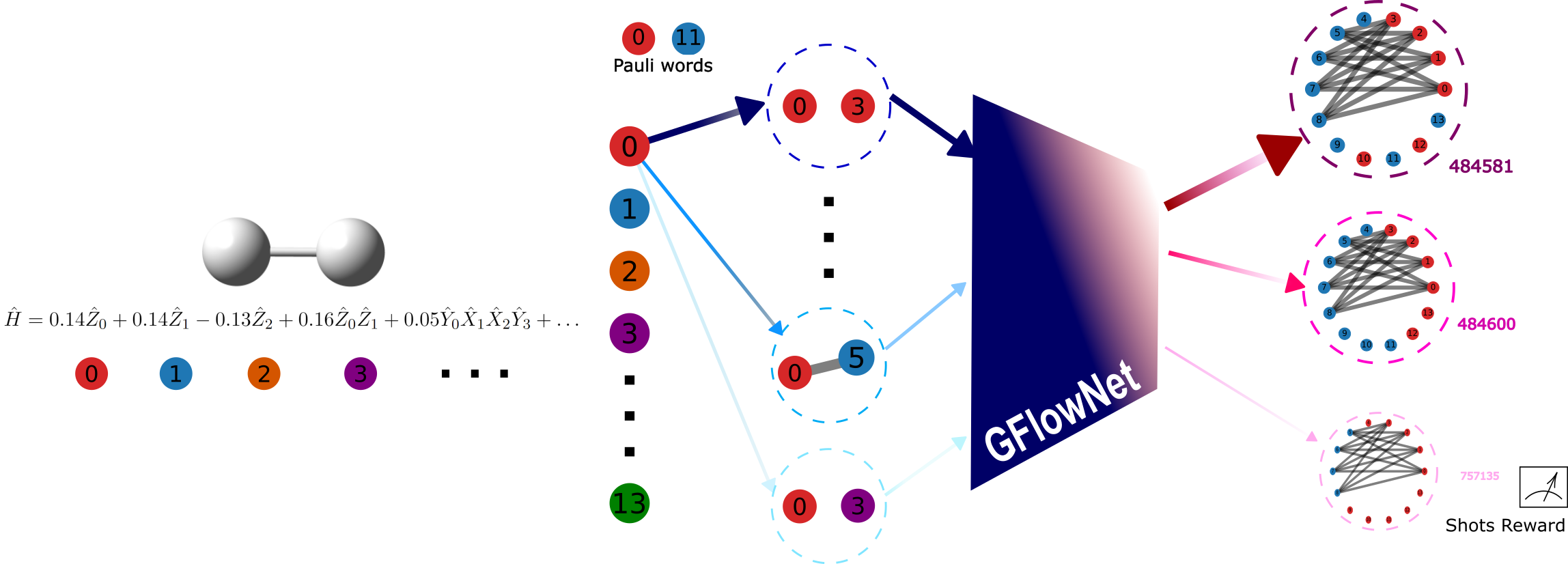}
    \caption{General scheme of GFlowNets for grouping Hamiltonian terms of H$_2$. Each node is a Pauli word from the Hamiltonian. The relative width of the arrows shows the probability of sampling a given state. The number of shots estimated to achieve chemical accuracy for each terminal state is shown in the last column.
    }
    \label{fig:diagram}
\end{figure}
In this work, we introduce the use of generative models for the measurement problem. Specifically, we utilize Generative Flow Networks (GFlowNets) to sample different valid groupings for a given Hamiltonian. GFlowNets are designed to sample from complex probability distributions by constructing flow-based policies to generate structured objects, such as graphs or sequences, through a series of intermediate steps \cite{bengio2021flow,bengio2023gflownet}. Unlike previous deterministic approaches, GFlowNets could learn a policy to sample the grouping configurations for a given Hamiltonian; see Fig.~\ref{fig:diagram}.
We build on the relationship between the graph representation of optimal groupings through the minimum-clique cover (MCC) and the coloring of its complementary graph ~\citep{izmaylov2020JCPmeasurements}. MCC is a known NP-hard problem with highly complex solution space which makes GFlowNets an attractive approach.
GFlowNets-based approaches have also been introduced for graph combinatorial problems~\citep{zhang2023let}, like molecule generation with target properties~\citep{bengio2021flow}. 
Due to their stochastic nature, GFlowNets offer a diverse set of high-quality solutions leading to a broad exploration, making them a promising approach for grouping strategies in tackling the measurement problem.

\section{Methods}
Here, we briefly introduce the GFlowNets algorithm, Section \ref{sec:gflownet}, and grouping term decomposition used for Hamiltonians, Section \ref{sec:lcu}. Finally, we give a description of our algorithm for the measurement problem.

\subsection{GFlowNets}\label{sec:gflownet}
The key feature of GFlowNets is the incremental generation of each object $x$ through a sequence of actions, which allows efficient sampling in complex, high-dimensional spaces \cite{bengio2021flow,bengio2023gflownet,zhang2023let,jain2023gflownets,zhang2022generative,malkin2022trajectory}. For a discrete set $\mathcal{X}$, the probability $P(x)$ to sequentially build $\mathcal{X}$ is given by
\begin{equation}
P(x) = \frac{R(x)}{Z}=\frac{R(x)}{\sum_{x^{\prime} \in \mathcal{X}} R\left(x^{\prime}\right)}.\label{eqn:boltzman_dist}
\end{equation}
where $R(x)$ is the reward function associated with state $x$, and $Z$ is the normalization constant. This formulation approximates the probability distribution $P(x)$ over the discrete set $\mathcal{X}$, proportional to the rewards of each state. Let $\mathcal{S}$ denote the set of states and $\mathcal{X} \subset \mathcal{S}$ denote the set of terminal states.

To sample from this probability distribution (Eq.~\ref{eqn:boltzman_dist}), we define a trajectory $\boldsymbol{s} = (s_0, s_1, \dots, s_n)$, where $s_0$ the initial state of any trajectory, progresses through intermediate states ($s_1\cdots s_{n-1}$) and ends with a valid terminal state ($s_n \in \mathcal{X}$). We define $\mathcal{T}$ as the set of all possible trajectories, and $\mathcal{T}_x$ as the subset of trajectories that terminate at a specific terminal state $x \in \mathcal{X}$. We then introduce a flow function $F : \mathcal{T} \rightarrow \mathbb{R}^+$, associated with a normalized probability distribution over trajectories, $P(\boldsymbol{s}) = {F(\boldsymbol{s})}/{Z}, \quad Z = \sum_{\boldsymbol{s} \in \mathcal{T}} F(\boldsymbol{s})$. A flow function is considered valid if, for each terminal state $x$, the total flow into $x$ matches its reward, $R(x) = \sum_{\boldsymbol{s} \in \mathcal{T}_x} F(\boldsymbol{s})$ \cite{bengio2021flow,bengio2023gflownet}. This implies that the probability $P(x)$, is proportional to its reward $P(x) = {\sum_{\boldsymbol{s} \in \mathcal{T}_x} F(\boldsymbol{s})}/{Z} \propto R(x)$.

If the trajectory is Markovian, $P(\boldsymbol{s})$ can be decomposed as $P\left(\boldsymbol{s}\right)=\prod_{i=1}^n P_F\left(s_i \mid s_{i-1}\right)$, where $P_F\left(s_i \mid s_{i-1}\right)$ is the forward policy corresponding to $F$. At each state, $P_F$ can be computed by $P_F\left(s^{\prime} \mid s\right)={F\left(s \rightarrow s^{\prime}\right)}/{F(s)}$, where $F(s) = \sum_{s'} F\left(s \rightarrow s^{\prime}\right)$. The flow matching constraint needs to be satisfied for all intermediate states \cite{bengio2021flow,bengio2023gflownet,malkin2022trajectory},
$\sum_{\left(s^{\prime \prime} \rightarrow s\right) \in \mathcal{A}} F\left(s^{\prime \prime} \rightarrow s\right)=\sum_{\left(s \rightarrow s^{\prime}\right) \in \mathcal{A}} F\left(s \rightarrow s^{\prime}\right)$,
where $\mathcal{A}$ denotes all possible transitions $(s \rightarrow s^{\prime})$.
Finally, we approximate the flow $F\left(s \rightarrow s^{\prime}\right)$ with a model $F_{\boldsymbol{\theta}}\left(s', s\right)$ with learnable parameter $\boldsymbol{\theta}$, which can be trained to minimize the loss $\mathcal{L}$ (Eq.~\ref{eqn:loss}) to satisfy the flow matching constraint.
{\footnotesize
\begin{equation}
\mathcal{L}(\boldsymbol{s})=\sum_{s\in \boldsymbol{s}, s \neq s_0}\left(\log \frac{\sum_{\left(s^{\prime \prime} \rightarrow s\right) \in \mathcal{A}} F_{\boldsymbol{\theta}}\left(s^{\prime \prime}, s\right)}{\sum_{\left(s \rightarrow s^{\prime}\right) \in \mathcal{A}} F_{\boldsymbol{\theta}}\left(s, s^{\prime}\right)}\right)^2 \label{eqn:loss}
\end{equation}}
For the terminal state $s_n$, it is important to note that the denominator becomes the reward $R(s_n)$.

\subsection{Hamiltonian terms grouping, measurements and GFlowNets implementation}\label{sec:lcu}
Variational Quantum Eigensolver (VQE) techniques for the ESP are one of the most common applications of quantum algorithms in NISQ devices. These approaches rely on measuring the expectation value of the Hamiltonian. The molecular Hamiltonian is defined by the identity of the atoms, interatomic distances, and the basis set chosen for the calculation. For its measurement, a fermion-to-qubit mapping is required, e.g., Jordan-Wigner (JW) and Bravyi-Kitaev~\citep{BK,JordanWigner}. We will focus on the JW mapping which takes the occupation number of orbitals and maps them directly to whether the qubit state is in $|0\rangle$ or $|1\rangle$ while maintaining the anticommutation relationships from fermionic operators by introducing a phase with $\hat{Z}$ qubit operators, leading to a qubit Hamiltonian in terms of Pauli products,
\begin{equation}
    \hat{H}_q=\sum_k^{N_P} c_k\hat{P}_k, \ \ \ \hat{P}_n = \bigotimes_{n = 1}^{N_q} \sigma_n, \label{eqn:qubitH}
\end{equation}
with every Pauli product $\hat{P}_n$ being a tensor product of Pauli operators and identities for the corresponding qubit $\sigma_n \in \{\hat{x}_n, \hat{y}_n, \hat{z}_n, \hat{1}_n \}$. $N_P$ is the number of Pauli words in $\hat{H}_{q}$. Specific details on this mapping can be found in \cite{JordanWigner, JWvsBK}.

To find molecular energies, VQE performs an iterative optimization of a parameterized wavefunction, $| \psi_\theta\rangle$, as $E_\theta = \min_\theta\langle\psi_\theta|\hat{H}_q|\psi_\theta\rangle$.
Since $\hat{H}_q$ contains terms that do not commute with each other, a single shot measuring of the operator is not possible. For this reason, we are required to partition the Hamiltonian into compatible fragments~\citep{izmaylov2020JCPmeasurements} and measure each of them separately
\begin{equation}
\hat{H}_q=\sum_{\alpha=1}^{N_f}\hat{H}_\alpha; \ \ \ \  E_\theta = \sum_{\alpha=1}^{N_f}\langle\psi_\theta|\hat{H}_\alpha|\psi_\theta\rangle.\label{eqn:VQE}
\end{equation}

Several grouping schemes exist, this work focuses on the fully commuting (FC) and qubit-wise commuting (QWC) groupings for the VQE problem. For FC, the requirement is that the Pauli products within a group commute with each other, $[\hat{P}_i,\hat{P}_j]=0$. QWC is a more strict condition since, for a pair of Pauli products, every single-qubit operator needs to commute $[\hat{P}_i,\hat{P}_j]_{QWC}=0$, e.g. the Pauli products $\hat{X}_1\hat{X}_2$ and $\hat{Y}_1\hat{Y}_2$ commute but not qubit-wise commute. Both of these partitionings can be achieved by constructing the corresponding commutativity graph and identifying the MCC. Each clique in the graph represents a distinct group. This is equivalent to coloring the complementary graph as is well-known in graph theory and as is often employed in quantum computing. Even though efficient algorithms are known for the coloring problem~\citep{Husfeldt_2015}, these do not guarantee that the resulting partitioning would be the best performing when implemented in quantum devices, an issue that we aim to tackle with GFlowNets. 

To sample optimal groupings using GFlowNets, we need to estimate the number of measurements ($M_{est}$) required to achieve a certain accuracy ($\varepsilon$). This quantity can be found as, $ M=\frac{1}{\varepsilon^2}\left(\sum_\alpha^{N_f}\sqrt{\text{Var}(\hat{H}_\alpha)}\right)^2\label{eqn:meas}$,~\citep{MeasFormula, Thomson-npj}
where $\text{Var}(\hat{H}_\alpha)$ is the variance of the operator which requires an approximation of the wavefunction or, ideally, the full-CI wavefunction to get the variances of each fragment. We approximate the variances under the assumptions, $\text{Cov}(\hat{P}_j,\hat{P}_k)=0$ and $\text{Var}(\hat{P}_j)\leq 1$, yielding the expression~\citep{MeasApprox}  
\begin{equation}
    M_{est}\approx \frac{1}{\varepsilon^2}\left( \sum_i\sqrt{\sum_j c_{ij}^2}\right). \label{eqn:meas_app}
\end{equation}
While it would be ideal to employ the exact wavefunction instead, this approximation is used in the reward function on the GFlowNets training stage due to its easiness of calculation. 

The reward employed is based on 1) the number of groupings (colors) on the graph and 2) the number of measurements required to achieve chemical accuracy of  1 kcal/mol or 1.6 mHa. Explicitly, we defined the reward function as,
\begin{equation}
    R(x)=\left( N_P-\text{max}\_\text{color}(x)\right) + \frac{\lambda_{0}}{M_{est}(x)}
\end{equation}
where $\text{max}\_\text{color}$ is the maximum color for the generated graph which gives the number of groups, the lower this number is, the fewer circuits we need to run.  $\lambda_0$ is a scaling factor, set to $10^6$ to account for the order of magnitude of the number of measurements. 

Now we proceed to describe our sampling algorithm which is schematized in Fig. \ref{fig:diagram}. First, a set of Pauli products from the molecular qubit Hamiltonian is fed to GFlowNets with the commutativity graph. The terms are assigned to a vertex, and edges are given by the commutation/QW-commutation relations to generate the complementary graph. At every training iteration, colors are assigned sequentially using a categorical distribution with probabilities dictated by Eq.~\ref{eqn:boltzman_dist}; a masking function is employed to limit the optimization space. The mask takes as the upper limit the number of groups from a classical heuristics-based algorithm, namely greedy coloring with a random sequential strategy, and it can be increased as the user requires. Each added color to the resulting graph represents a different state, $x$, through the action sequence for GFlowNets whose transitions come from adding a new colored term to the list. Once a terminal state is reached, the validity of the graph coloring is assessed, $R(x)$ is calculated, and the color probabilities per vertex given by GFlowNets are updated. Finally, the total number of colors in a terminal state is the number of groups to be measured and a new iteration begins.

\section{Results}
In this section, we present the results of applying GFlowNets to six molecular systems: \ce{H2}, \ce{H4}, \ce{LiH}, \ce{BH}, \ce{BeH2}, and \ce{N2}. For all systems, we used an inter-atomic distance of 1 Å, the STO-3G basis set~\citep{STO-3G}, and the Jordan-Wigner mapping. Except for \ce{N2}, where only 100 training iterations were used, 1,000 iterations were found sufficient for the other systems. The number of Pauli terms ($N_{P}$) are reported in Table~\ref{tab:measurements}.
For $F_{\boldsymbol{\theta}}(s',s)$, we employed a two-layer MLP with 512 neurons and the $\tanh$ activation function. For all simulations, we used an NVIDIA GeForce RTX 4080 SUPER GPU, Torch, the Adam optimizer~\citep{adam} with a learning rate of $3\times10^{-4}$ and gradient accumulation every ten steps. 

\begin{table}[ht!]
    
\caption{Comparison of $M_{est}$ (in millions) required to achieve chemical accuracy for the different available methods. Number of colors reported in parentheses. Full is the $M_{est}$ without grouping. We define the Reduction factor as the ratio between GFlowNets and the best-performing greedy NetworkX-method. 
    }
    \centering
\scriptsize
    \begin{tabular}{|c|c|c|c|c|c|c|c|c|c|} \hline  
        &   \multicolumn{2}{|c|}{GFlowNets (ours)}&  \multicolumn{2}{|c|}{NX-lf}&  \multicolumn{2}{|c|}{NX-dsat} &Full&\multicolumn{2}{|c|}{Reduction factor}\\ \hline  
 System [$N_{P}$]& FC& QWC& FC& QWC& FC& QWC& None& FC&QWC\\ \hline  
 H$_2$ [14]&  \textbf{0.485 (2)}& \textbf{ 0.746 (5)}&  0.757 (2)&  \textbf{0.746 (5)}&  0.757 (2)&  \textbf{0.746 (5)}& 2.56& 0.640&1.000\\ \hline  
 H$_4$ [184]& \textbf{1.50 (19)}& \textbf{1.13 (71)}& 2.87 (11)& 1.29 (68)& 3.79 (9)& 1.27 (67)& 76.7& 0.524&0.877\\ \hline  
 LiH [275]&  \textbf{2.10 (23)}&  \textbf{4.10 (72)}&  4.13 (21)&  6.77 (64)&  8.32 (12)&  6.54 (63)& 121& 0.508&0.626\\ \hline  
 BH [275]&  \textbf{3.23 (22)}&  \textbf{4.92 (72)}&  5.99 (21)&  8.21 (64)&  10.5 (12)&  7.78 (63)& 610& 0.539&0.632\\ \hline  
 BeH$_2$ [326]& \textbf{ 5.71 (23)}&  \textbf{9.78 (101)}&  14.0 (16)&  21.0 (100)&  22.4 (9)&  21.1 (99)& 405& 0.407&0.465\\ \hline  
N$_2^*$[824]& \textbf{22.8 (49)}& \textbf{48.0 (340)}& 51.2 (23)& 111 (318)& 125 (16)& 110 (314)& 12705& 0.446&0.433\\ \hline 
    \end{tabular}

    \label{tab:measurements}
\end{table}

Table \ref{tab:measurements} compares the estimated required number of measurements for the groupings obtained by GFlowNets and the ones from greedy coloring in NetworkX following the Largest first (NX-lf), and saturation largest first (NX-dsat) strategies. 
We found that to achieve chemical accuracy, GFlowNets on average require 51\% for FC and 67\% for QWC, of measurements compared to heuristic methods. These promising results encourage further developments for incorporation of state-of-the-art techniques, such as $k$-commutativity groupings~\citep{k-commutativity} and ghost Pauli products~\citep{Choi2022} to the GFlowNets workflow to further improve it.

\begin{figure}[ht!]
    \centering
    \includegraphics[scale=0.385]{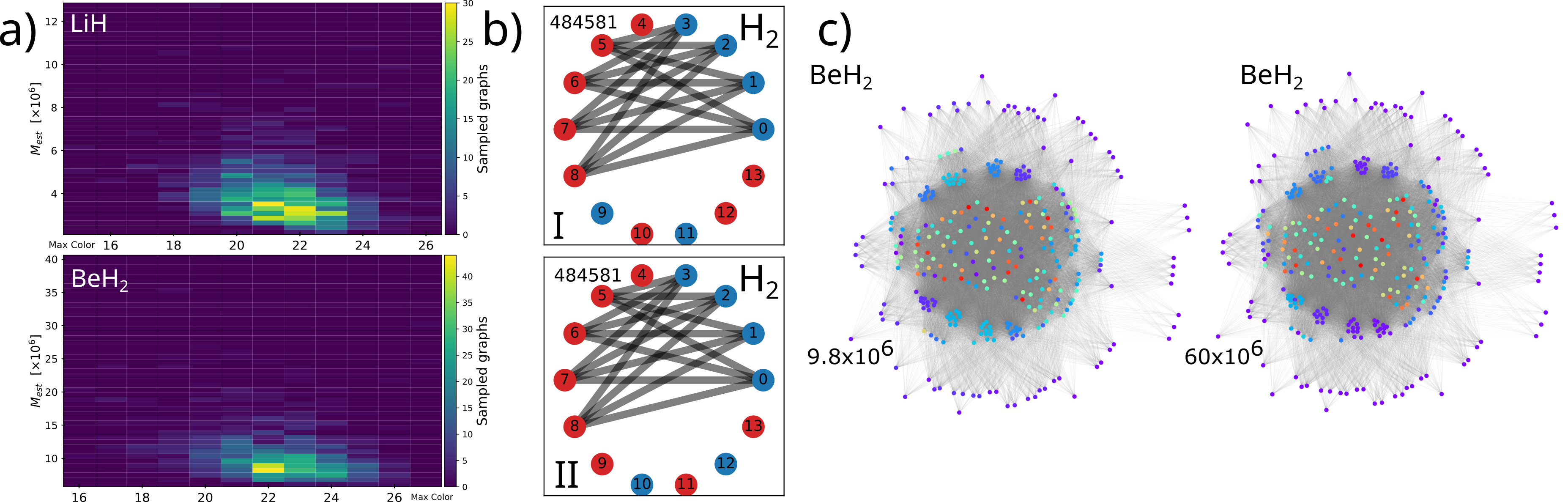}
    \caption{a) 2D histogram of number of groupings and estimated measurements ($M_{est}$) for \ce{LiH} and \ce{BeH2} with FC grouping. b) Equivalent graphs sampled for \ce{H2} in FC grouping. c) Sampled graphs for \ce{BeH2} in QWC grouping. $M_{est}$ shown for each graph.}
    \label{fig:results}
\end{figure}

In Fig.~\ref{fig:results}a, we show the 2D histogram of the maximum color found and the estimated number of measurements $M_{est}$ for \ce{LiH} and \ce{BeH2} in the FC grouping. Let's reiterate that the maximum color is the number of compatible groups. It is appreciable how the sampling algorithm veers towards configurations with a balance between the number of groups and $M_{est}$. This effect is of course dependent on  $\lambda_0$ and optimization of this parameter is one of the perspectives of the present work. Decreasing this parameter, samples colorings with a lower number of groups while the opposite biases the algorithm towards lower $M_{est}$ regardless of the group number. For future work, we plan to extend our algorithm to incorporate classically efficient wavefunctions, such as Coupled-Cluster, to improve variance estimation~\citep{Thomson-npj} thereby providing a more accurate reflection of our algorithm's practical performance.

Some of the sampled graphs generated by GFlowNets for \ce{H2} and \ce{BeH2} Hamiltonians are shown in Fig.~\ref{fig:results}b-~\ref{fig:results}c. As we observe, \ce{H2} graphs I and II are equivalent colorings, yielding the same number of measurements. The set of equivalent graphs is responsible for the increased number of sampled graphs shown in the histograms for a particular $M_{est}$. On the other hand, the \ce{BeH2} graphs show how inefficient groupings can increase dramatically the required number of measurements, producing a sixfold increase in $M_{est}$.

It is important to note that incorporating the number of measurements into the reward function, allows the algorithm to sample partitionings that may not minimize the number of cliques but result in lower measurement estimates. This flexibility in the choice of GFlowNets' reward function opens research opportunities for its implementation towards other resource optimization problems in quantum computing. Our results provide early evidence that GFlowNets, coupled with more competitive techniques, could become a promising alternative for grouping compatible operators. 

\section{Summary} 
In this work, we introduced GFlowNets for grouping compatible Hamiltonian terms in VQEs. 
Our results demonstrate that GFlowNets can achieve more efficient groupings compared to the deterministic methods available in NetworkX, as used in Pennylane~\citep{pennylane}. This research serves as a promising foundation for leveraging GFlowNets to address the measurement problem. Given the flexibility of GFlowNets' reward function, potential extensions to our framework include exploring $k$-commutativity conditions~\citep{k-commutativity}, incorporating ``ghost'' Pauli products~\citep{Choi2022} or utilizing the Majorana tensor representation~\citep{NachoMajorana} for more general decompositions that further reduce measurement requirements. These are all directions we plan to investigate in the near future.
\section*{Acknowledgments}
The authors thank Ignacio Loaiza for fruitful discussions. 
This research was partly enabled by support from the Digital Research Alliance of Canada and NSERC Discovery Grant No. RGPIN-2024-06594. GR acknowledges the support of CIFAR through the CCAI chair program. 

\newpage
\bibliographystyle{plainnat-sortby-first}
\bibliography{references}



\end{document}